\documentclass[english,journal=jacsat,manuscript=article,etalmode=truncate,maxauthors=0,layout=twocolumn,number]{achemso}
\setkeys{acs}{etalmode=truncate,maxauthors=0,articletitle=true,email=true}

\usepackage{amsmath}             %
\usepackage{mathtools}             %
\usepackage{amssymb}             %
\usepackage{wasysym}             %
\usepackage{color}               %
\usepackage{bm}
\usepackage{setspace}            %
\usepackage{graphicx}            %
\usepackage{diagbox}
\usepackage{placeins}
\usepackage{xurl}
\usepackage{caption}
\usepackage[colorlinks=true,allcolors=blue]{hyperref}
\usepackage{wrapfig}             %
\usepackage[dvipsnames]{xcolor}  %
\usepackage{array,booktabs,multirow}      %
\usepackage{comment}
\usepackage[export]{adjustbox}
\usepackage{arydshln}
\usepackage{multirow, multicol}
\usepackage[ruled, vlined, linesnumbered]{algorithm2e}

\SectionNumbersOn    %

\usepackage[font=footnotesize,labelfont=bf,labelsep=period,width=0.95\textwidth]{caption}   %
\usepackage[font=scriptsize,labelfont=bf]{subcaption}    
\usepackage{hyperref}
\usepackage[capitalize]{cleveref}
\crefname{figure}{Figure}{Figures}
\Crefname{figure}{Figure}{Figures}
\crefname{table}{Tab.}{Tabs.}
\Crefname{table}{Table}{Tables}
\crefname{equation}{Eq.}{Eqs.}
\Crefname{equation}{Eq.}{Eqs.}
\crefname{section}{Sec.}{Secs.}
\Crefname{section}{Section}{Sections}

\title{Electronic-Structure-Informed Many-Body Encoding for Quantum Machine Learning}

\author{Liangnuo Zhang}
\affiliation[University of Washington]
{Department of Chemistry, University of Washington, Seattle, WA 98195, USA}

\author{Tanner Culpitt}
\affiliation[University of Washington]
{Department of Chemistry, University of Washington, Seattle, WA 98195, USA}
\email{tculpitt@uw.edu}

\author{Qiwei Yin}
\affiliation[University of Washington]
{Department of Chemistry, University of Washington, Seattle, WA 98195, USA}

\author{Linghua Zhu}
\affiliation[University of Washington]
{Department of Chemistry, University of Washington, Seattle, WA 98195, USA}
\alsoaffiliation[Colorado School of Mines]
{Department of Computer Science, Colorado School of Mines, Golden, CO 80401, USA}
\email{linghua.zhu@mines.edu}

\author{Xiaosong Li}
\affiliation[University of Washington]
{Department of Chemistry, University of Washington, Seattle, WA 98195, USA}
\email{xsli@uw.edu}

\begin{document}

\twocolumn[
\begin{@twocolumnfalse}
\begin{abstract}
Data encoding is a central challenge in quantum machine learning (QML), as it determines how classical information is mapped onto quantum states and influences the expressive power and efficiency of parameterized quantum circuits. Here, we introduce a physics-informed quantum encoding strategy that exploits the intrinsic many-body structure of electronic systems to establish a direct correspondence between the body order of input descriptors and the locality of quantum gates. Specifically, $n$-body electronic descriptors obtained from a single mean-field electronic-structure calculation are encoded through $n$-qubit gates, yielding a systematic hierarchy of quantum circuits with increasing many-body expressivity. Using reduced-density-matrix descriptors, we construct quantum machine-learning models incorporating one- through four-body encodings and benchmark their performance in predicting electron-correlation energies along bond-stretching potential energy surfaces. We demonstrate that increasing the body order of the encoding systematically improves prediction accuracy toward the full configuration interaction limit. Remarkably, disconnected many-particle densities constructed from lower-order electronic information achieve accuracy comparable to explicitly correlated two-particle densities in key cases, providing a computationally inexpensive route to higher-body quantum encodings. These findings establish a connection between the many-body structure of classical data and the design of quantum feature maps, offering a systematic, physically interpretable framework for enhancing the expressivity and predictive performance of quantum machine-learning circuits.
\end{abstract}
\end{@twocolumnfalse}
]

\section{Introduction}
\label{sec:intro}

Quantum machine learning (QML) has emerged as a promising route to predicting molecular properties \cite{Lloyd17_195, Coles21_625, Kais22_6475,Li26_030901}, with the hope that
quantum circuits can represent correlations in chemical data that are hard to capture classically \cite{McClean21_2631}. A central question in this program is how a molecule should be encoded into a circuit\cite{Killoran19_040504, Sweke21_582}. The encoding decides what information the state carries. Training starts after that. Whatever it leaves out, the trainable part of the circuit cannot recover. On current hardware, where circuits are shallow and gates are noisy \cite{Preskill18_79}, a feature map that spreads information thinly leaves little room for the model to do useful work~\cite{Wild22_042407}. The encoding is therefore not a preprocessing detail. It is where the physics of the problem either enters the computation or not.

\begin{figure*}[ht!]
\centering
\includegraphics[width=\linewidth]{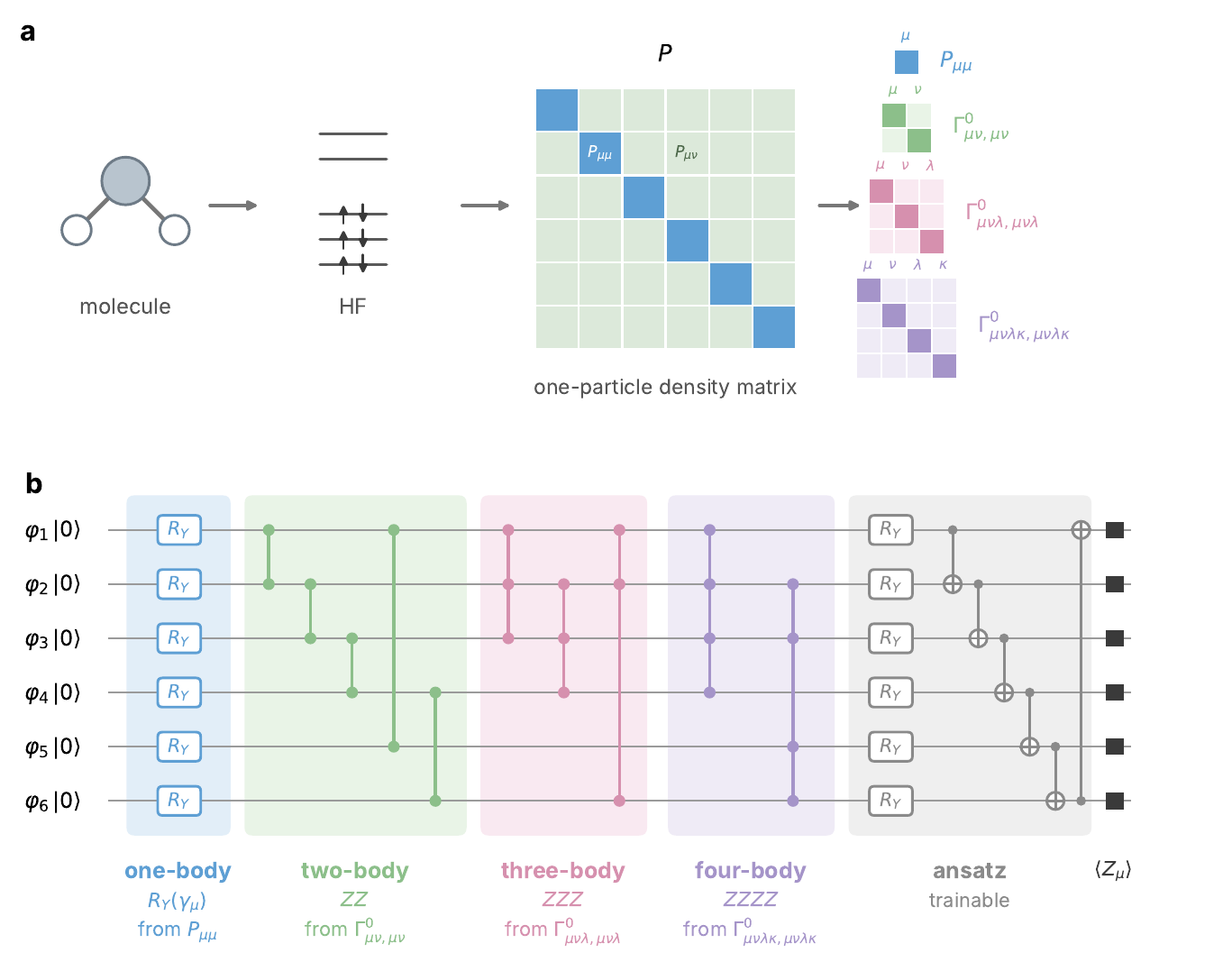}
\caption{{\bf Electronic-Structure-Informed Quantum Encoding}. \textbf{(a)}, A single Hartree–Fock (HF) calculation provides the one-particle density matrix, $\mathbf{P}$, in the L\"owdin-orthogonalized atomic orbital basis. Electronic-structure descriptors are extracted from the diagonal elements of the one-particle reduced density matrix (RDM), $P_{\mu\mu}$, and the disconnected two-, three-, and four-particle RDMs, $\Gamma^{0}_{\mu\nu,\mu\nu}$, $\Gamma^{0}_{\mu\nu\lambda,\mu\nu\lambda}$, and $\Gamma^{0}_{\mu\nu\lambda\kappa,\mu\nu\lambda\kappa}$, respectively. \textbf{(b)}, Each L\"owdin orbital, $\varphi{\mu}$, is mapped onto a single qubit. The one-particle descriptors are encoded through single-qubit $R_Y(\gamma_{\mu})$ rotations, whereas the higher-order RDM descriptors are encoded through two-, three-, and four-qubit $ZZ$, $ZZZ$, and $ZZZZ$ rotations acting on the corresponding orbital qubits. The resulting electronic-structure-informed quantum state is further transformed by a shallow parameterized ansatz, followed by measurements of $\langle Z_{\mu}\rangle$ to generate the output features.}
\label{fig:concept}
\end{figure*}

Most encoding schemes treat the molecule as a generic feature vector. Fingerprint and angle encodings map descriptors onto rotation angles without reference to the underlying chemistry \cite{Gambetta19_209, Meyer21_032430, Latorre20_226}. Work on inductive bias points the other way, models that carry little knowledge of the problem train and generalize poorly~\cite{Cerezo22_030341}, and a quantum model can be expected to help only when such knowledge is built into the circuit~\cite{Scholkopf21_12661}. Whether a variational model succeeds depends on whether its parametrization carries the right inductive bias, yet there are few guiding principles for building one~\cite{Eisert23_010328,Schuld24_07059}. Geometric QML gives one, building circuits that respect the symmetry group of the data~\cite{Cerezo22_030341,Eisert23_010328,Cerezo22_07980,Cerezo24_020328}. Quantum graph neural networks give another one, letting the circuit follow the connectivity of a graph~\cite{Hidary19_12264}. The $ZZ$ feature map does place pairwise terms on two-qubit gates~\cite{Gambetta19_209}. But these terms are products of single features, so they add little information beyond what the single-qubit layer already holds. 

A more physical approach is to build the encoding from the structure of the molecule itself. The quantum molecular structure encoding (QMSE) scheme does this by writing bond orders and interatomic couplings, collected in a hybrid Coulomb-adjacency matrix, into one- and two-qubit rotations \cite{Wales25_045076}. This is a clear improvement over generic feature maps, because the circuit now reflects the connectivity of the molecule. Pegg \textit{et al.} follow the same idea, mapping atoms to qubits and letting bonds decide which pairs interact~\cite{Wu26_13737}. A classical model with the same structure does about as well, and the authors trace the gain to the topology-aligned inductive bias. But what these encodings carry is the molecular graph, not the electronic structure. The encoding sees nuclear charges, distances, and bonds, and it stops at two-body terms. What the circuit never sees is the electronic response that distinguishes molecules of similar geometry. A different route feeds the molecular ground state itself into a symmetry-invariant quantum neural network, which predicts excited-state properties across geometries from few training points~\cite{Roth25_09423}. 
Electronic-structure data have also served as inputs to quantum neural networks representing exchange-correlation functionals,\cite{Elfving26_012427} and hybrid quantum--classical models of proton affinities have drawn their inputs from a descriptor set that includes quantum-chemical quantities.\cite{Merz25_2235} Quantum circuits have also been used to prepare multiconfigurational molecular wavefunctions\cite{Yanai23_634} and to estimate electron-correlation energies.\cite{Angelakis24_025003}

Electronic information is known to matter. In classical machine learning, the spectrum of approximated Hamiltonian matrices (SPAHM) family of representations is built from one-electron Hamiltonians and density matrices rather than from geometry \cite{Corminboeuf22_286, Corminboeuf24_1108}. Such electronic-structure-based representations outperform purely structural ones in cases where nuclear coordinates do not uniquely characterize the electronic state, including charged and open-shell species \cite{Corminboeuf24_1108}. Correlation energies have likewise been learned from Hartree--Fock molecular-orbital features~\cite{Miller18_4772} and the density matrices.~\cite{E20_7155} More broadly, machine learning built on electronic-structure quantities is a well-developed area, spanning $\Delta$-machine learning,\cite{vonLilienfeld15_2087} further molecular-orbital and atomic-orbital based models,\cite{Miller19_131103,Miller21_064108,Miller20_124111,vonLilienfeld22_114101} models built on MP2-level quantities such as amplitudes, energy components
and densities,\cite{Reuter18_6343,Vogiatzis19_4129,Vogiatzis20_7453,Shaw17_161725,Crawford20_4861} and machine-learned functionals and density-matrix methods.\cite{FernandezSerra20_3509,Pavanello23_6281} The key point is that quantities such as the density matrix contain information about the electronic state that is absent from the nuclear coordinates.

Here we bring this principle directly to quantum circuits (\cref{fig:concept}). Rather than reducing the molecule to a generic feature vector, we exploit the correspondence between the body order of electronic-structure descriptors and the locality of quantum gates. A $k$-body descriptor involves $k$ orbitals, while a $k$-qubit gate acts on $k$ qubits; the descriptor can therefore parameterize a gate of the same locality directly. We assign one qubit to each orbital and obtain all descriptors from a single electronic-structure calculation.

One-body quantities determine single-qubit rotations, with each qubit parameterized by the population of its associated orbital. Higher-body quantities determine multi-qubit entangling gates, allowing the circuit connectivity and interaction strengths to reflect the underlying electronic structure. In particular, descriptors derived from the $n$-particle density matrix parameterize $n$-qubit gates, so that the many-orbital structure of the Hartree--Fock state is placed explicitly on the gates that act on those orbitals. No fermion-to-qubit transformation or Pauli decomposition is required: electronic-structure quantities enter directly as numerical gate parameters rather than as operators that must be mapped to or measured on the quantum device.

This design has three key advantages. First, the encoding represents the electronic state rather than geometry alone. Because the density matrices already reflect the electronic response to the nuclear framework, the circuit begins from a chemically informed representation that a shallow trainable ansatz can map to the target property. Second, interactions between weakly coupled or distant orbitals can be omitted with minimal loss of information, allowing the circuit topology to follow the molecular electronic structure rather than a fixed connectivity pattern. Third, shared functions map orbital descriptors to gate angles, and variational parameters are shared across qubits. The number of encoding and ansatz parameters therefore does not grow with the molecular size. 

This work makes three main contributions. First, we introduce a many-body encoding scheme based on disconnected many-particle reduced density matrices, which can be constructed directly from lower-order electronic-structure information. Second, we benchmark the resulting QML models for recovering electron-correlation energies toward the full configuration interaction limit, comparing their performance with a geometry-based model across bond-stretching potential energy surfaces for five molecules. Third, we provide a physical interpretation of the hierarchical encoding layers by identifying the electronic-structure information carried by each layer and elucidating how that information contributes to learning electron correlation.

\begin{table*}[ht!]
\captionsetup{width=\linewidth}
\footnotesize
\caption{MB$n$ Models. The rotation angle is defined as $\gamma = \frac{\pi}{2}\tanh(\mathrm{MB}n)$. The MB1 information is encoded using a single-qubit $R_Y(\gamma_\mu)$ rotation, whereas higher-order MB$n$ information ($n \geq 2$) is encoded through multi-qubit rotations of the form $\exp\left(-\frac{i}{2}\gamma_{\mu\nu\cdots}Z_\mu Z_\nu\cdots\right)$. The full expressions for the MB3 and MB4 models can be found in \cref{eq:gamma03} and \cref{eq:gamma04}, respectively.}
\label{tab:MBn}
\centering
\begin{tabular}{c|c|c|c}
\toprule
MB1 & MB2 &  MB2$^{\Gamma^0}$ & MB2$^\Gamma$\\
\midrule
$a_{1}\tilde{P}_{\mu\mu}$ &$a_2 \tilde{P}_{\mu\mu}\tilde{P}_{\nu\nu}+b_2 \tilde{P}_{\mu\nu}\tilde{P}_{\mu\nu}$&$a_2\tilde\Gamma^0_{\mu\nu,\mu\nu}$&$a_2\tilde\Gamma_{\mu\nu,\mu\nu}$\\
\bottomrule
\end{tabular}

\vspace{10pt}
\begin{tabular}{c|c|c|c}
\toprule
MB3 &MB3$^{\Gamma^0}$& MB4&MB4$^{\Gamma^0}$\\
\midrule
$a_3 \tilde{P}_{\mu\mu}\tilde{P}_{\nu\nu}\tilde{P}_{\lambda\lambda}$&$a_3 \tilde\Gamma^0_{\mu\nu\lambda,\mu\nu\lambda}$&
$a_4 \tilde{P}_{\mu\mu}\tilde{P}_{\nu\nu}\tilde{P}_{\lambda\lambda}\tilde{P}_{\kappa\kappa}$&$a_4 \tilde\Gamma^0_{\mu\nu\lambda\kappa,\mu\nu\lambda\kappa}$\\
$+b_3(\tilde{P}_{\mu\mu}\tilde{P}_{\nu\lambda}\tilde{P}_{\lambda\nu}$+...)&&
$+b_4(\tilde{P}_{\mu\mu}\tilde{P}_{\nu\nu}\tilde{P}_{\lambda\kappa}\tilde{P}_{\kappa\lambda}$+...)&\\
$+c_3(\tilde{P}_{\mu\nu}\tilde{P}_{\mu\lambda}\tilde{P}_{\nu\lambda}$+...)&&
$+c_4(\tilde{P}_{\kappa\lambda}\tilde{P}_{\lambda\kappa}\tilde{P}_{\mu\nu}\tilde{P}_{\nu\mu}$+...)&\\
&&$+d_4(\tilde{P}_{\mu\nu}\tilde{P}_{\mu\lambda}\tilde{P}_{\nu\lambda}\tilde{P}_{\kappa\kappa}$+...)&\\
&&$+e_4(\tilde{P}_{\mu\nu}\tilde{P}_{\mu\lambda}\tilde{P}_{\nu\kappa}\tilde{P}_{\lambda\kappa}$+...)&\\
\bottomrule
\end{tabular}
\end{table*}

\section{Results}
\label{sec:results}

\subsection{Training Target}
The QML model inputs consist of mean-field quantities in the many-body encoding scheme  mapped onto one to four quantum gates (see \cref{sec:many-body}), whereas
the learning target is the correlation energy. The target quantity is the full configuration interaction (FCI) correlation energy relative to the Hartree--Fock reference,
\begin{equation}
  E_{\mathrm{corr}}
  =
  E_{\mathrm{FCI}}
  -
  E_{\mathrm{HF}},
  \label{eq:correlation_energy_target}
\end{equation}
where $E_{\mathrm{FCI}}$ and $E_{\mathrm{HF}}$ are evaluated using the same basis (see \cref{sec:abintio} for computational details). Characteristics of the training dataset are presented in \cref{sec:molecules} and \cref{fig:PES}.

\subsection{Systematic Convergence of the Many-Body Encoding Hierarchy}
\label{sec:scan}

We first examine the hierarchy of many-body (MB$n$) models. \Cref{tab:MBn} summarizes the descriptors employed at different levels of many-body encoding, the variants considered in this section, and their corresponding quantum-gate implementations. A detailed discussion of the descriptor choices and their construction is provided in \cref{sec:mb_descriptor}, while their incorporation into the QML model through $n$-qubit gates is described in \cref{sec:many-body}.

MB1 includes only the one-body layer; MB2 augments MB1 with the two-body layer; MB3 further adds the three-body layer; and MB4 includes all one- through four-body layers. Each qubit represents one L\"owdin-orthogonalized atomic orbital, resulting in registers of five to ten qubits for the molecules studied here. However, the circuit complexity increases substantially from MB1 to MB4, as reflected in the growing number of quantum gates.

\begin{figure}[ht!]
\centering
\includegraphics[width=\linewidth]{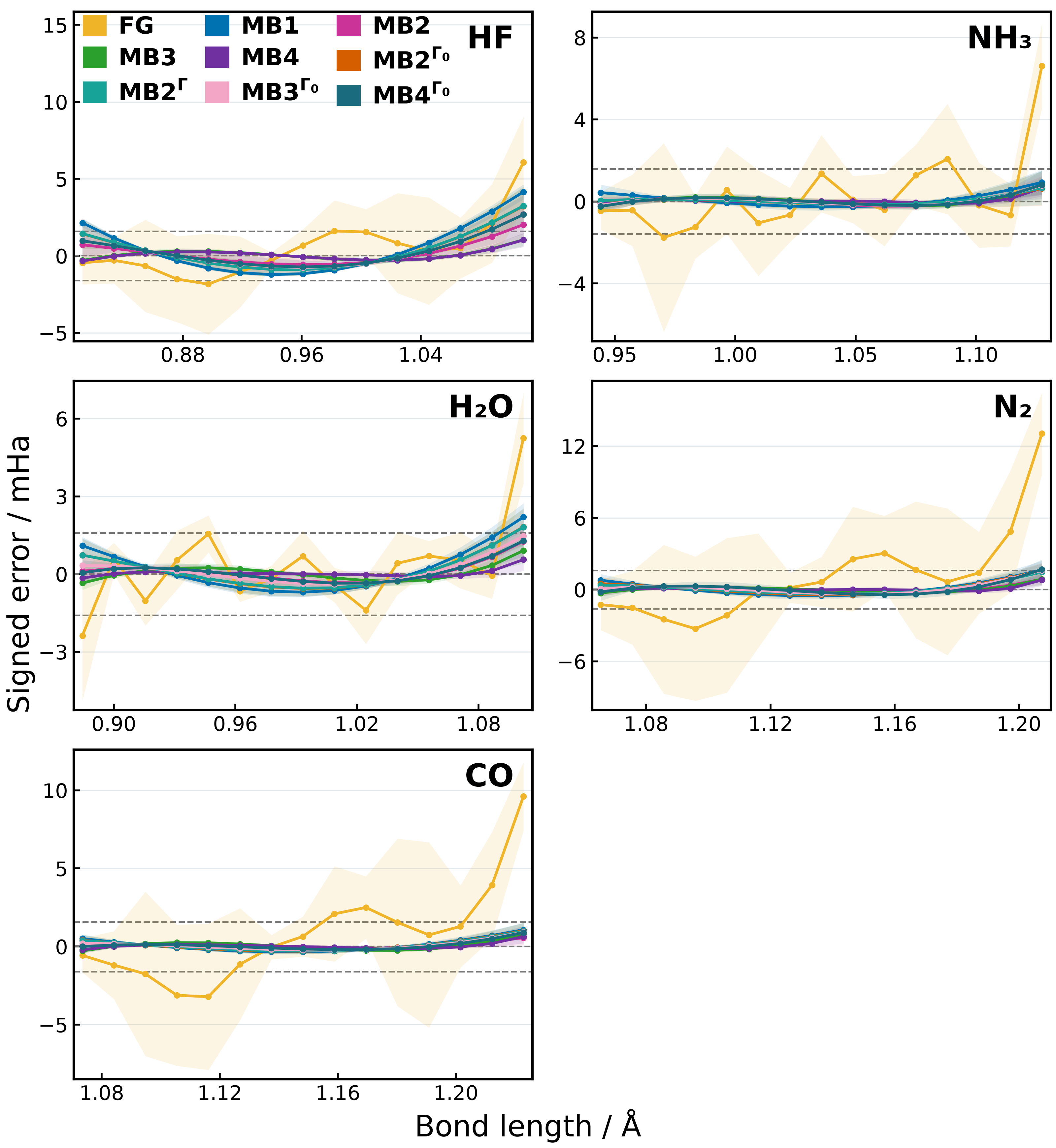}
\captionsetup{width=\linewidth}
\caption{Mean signed error along the bond-stretching scans. Signed errors of the FG, MB1, MB2, MB3, MB4, MB2$^\Gamma$, MB2$^{\Gamma^0}$, MB3$^{\Gamma^0}$, and MB4$^{\Gamma^0}$ models as functions of bond length for all five molecules. Solid lines show the mean over 64 random seeds, and shaded bands indicate standard error. Dashed horizontal lines mark the chemical-accuracy threshold of $\pm 1.6$ mHartree.}
\label{fig:convergence}
\end{figure}

\begin{figure}[ht]
\centering
\includegraphics[width=\linewidth]{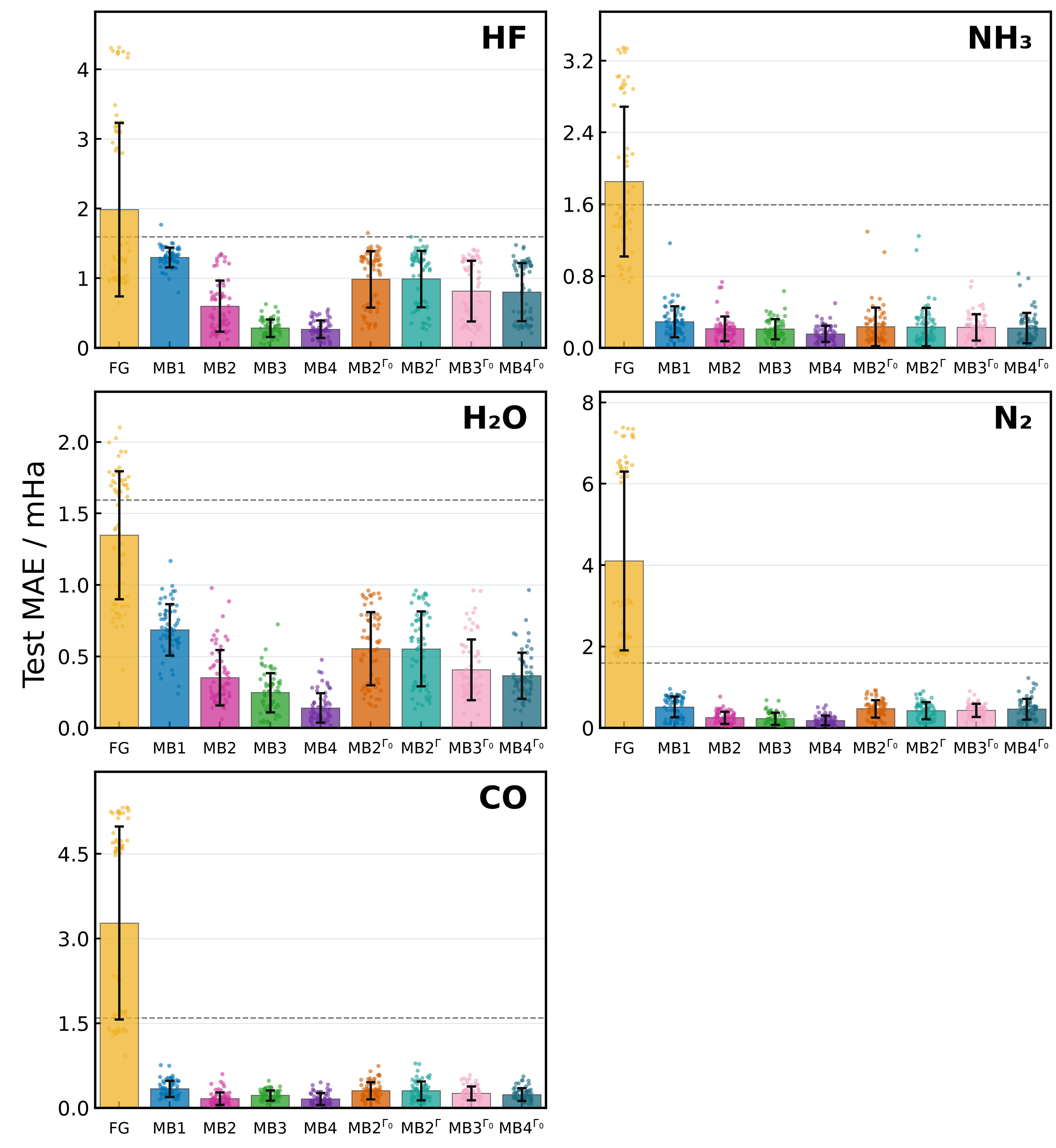}
\captionsetup{width=\linewidth}
\caption{Mean absolute errors of the FG, MB1, MB2, MB3, MB4, MB2$^\Gamma$, MB2$^{\Gamma^0}$, MB3$^{\Gamma^0}$, and MB4$^{\Gamma^0}$ models for all five test molecules. The error bar indicates standard error. Dashed horizontal lines mark the chemical-accuracy threshold of $\pm 1.6$ mHartree.}
\label{fig:comparison}
\end{figure}

\Cref{fig:convergence} shows the
signed error against bond length for all five test molecules. The error is signed rather than absolute, so the plots show where a model is biased and in which direction, not only how large the error is.

All MB models substantially outperform the geometric fingerprint (FG) approach, with a clear advantage already evident at the MB1 level. The geometric fingerprint alone does not adequately capture how electronic correlation evolves with molecular geometry, leading to larger errors and less systematic predictive behavior along the bond-stretching coordinate.
In contrast, the MB$n$ models directly encode electronic-structure information onto the quantum hardware, allowing the model to correlate fluctuations in the underlying electronic structure with changes along the potential energy scan. This direct access to electronic-structure descriptors provides a more physically informed representation of the evolving molecular state and leads to more systematic and accurate predictions.

\Cref{fig:comparison} and \Cref{tab:maeall} show the overall statistics over the test geometries. As higher-order electronic-structure information is incorporated, the QML models show systematic improvement. For most test molecules, MB1 already predicts the electron correlation energy within chemical accuracy (1 kcal/mol), while MB2-MB4 further improve the predictions. Importantly, the improvement is reflected not only in the mean error but also in the reduced variability across different random seeds, as indicated by the progressively smaller standard error. At the MB4 level, all test cases achieve correlation-energy prediction errors below the chemical-accuracy threshold.

\begin{table*}[t]
\captionsetup{width=\linewidth}
\caption{Test mean absolute errors for all five molecules. Means with standard errors of the mean in parentheses over 64 seeds, in mHartree. Molecules are ordered by the number of qubits.}
\label{tab:maeall}
\centering
\footnotesize
\setlength{\tabcolsep}{5pt}
\begin{tabular}{@{}lccccccccc@{}}
\toprule
 & FG & MB1 & MB2 & MB3 & MB4 & MB2$^{\Gamma^0}$ & MB2$^\Gamma$ & MB3$^{\Gamma_0}$ & MB4$^{\Gamma_0}$ \\
\midrule
HF & 1.99(0.16) & 1.30(0.02) & 0.60(0.05) & 0.28(0.02) & 0.27(0.02) & 0.98(0.05) & 0.99(0.05) & 0.82(0.05) & 0.80(0.05) \\
H$_2$O & 1.35(0.06) & 0.69(0.02) & 0.35(0.02) & 0.25(0.02) & 0.14(0.01) & 0.55(0.03) & 0.55(0.03) & 0.41(0.03) & 0.37(0.02) \\
NH$_3$ & 1.85(0.10) & 0.29(0.02) & 0.21(0.02) & 0.21(0.01) & 0.16(0.01) & 0.24(0.03) & 0.23(0.03) & 0.23(0.02) & 0.22(0.02) \\
N$_2$ & 4.10(0.27) & 0.52(0.03) & 0.25(0.02) & 0.23(0.02) & 0.18(0.01) & 0.47(0.03) & 0.43(0.03) & 0.43(0.02) & 0.46(0.03) \\
CO & 3.28(0.21) & 0.34(0.02) & 0.16(0.01) & 0.22(0.01) & 0.16(0.01) & 0.31(0.02) & 0.30(0.02) & 0.26(0.02) & 0.24(0.01) \\
\bottomrule
\end{tabular}
\end{table*}

\subsection{Hierarchical Contributions of Many-Body Information}

\subsubsection{One-body layer}

The analyses in the previous section suggest that molecular electronic-structure information is more effectively represented by the many-body encoding scheme. In contrast, the geometric fingerprint (FG) approach encodes only structural descriptors, including interatomic distances, inverse distances, and bond angles. From the perspective of electronic-structure theory, such geometric information alone is insufficient to fully capture the changes in electron correlation that are missing from the underlying mean-field description.

The MB1 model (\cref{sec:MB1}) uses the diagonal elements of the density matrix in the orthonormalized atomic-orbital basis, which encode the electron occupations associated with individual atomic orbitals. These density-matrix quantities are also fundamental ingredients in correlated electronic-structure methods for recovering electron correlation energy. Therefore, the diagonal elements provide the QML model with direct information about how the electronic population redistributes as the molecular geometry changes. Although MB1 contains only one-body information, it already captures physically meaningful variations in the electronic structure that are absent from purely geometric descriptors, explaining its substantial improvement over the FG model.

\subsubsection{Two-body layer}

The two-body layer (\cref{sec:MB2}) encodes pairwise electronic information using descriptors derived from the disconnected two-particle reduced density matrix (RDM) defined in \cref{eq:gamma02}. Consequently, the entangling structure of the circuit reflects the pairwise electronic couplings captured by this two-particle density information. Because the full two-particle RDM is a four-index tensor, it cannot be mapped directly onto a two-qubit gate. We therefore restrict the two-body encoding to the diagonal elements of the two-particle RDM.

The first two-body encoding model introduced here is based on the standardized disconnected two-particle RDM, $\tilde\Gamma^0_{\mu\nu,\mu\nu}$, in \cref{eq:gamma02}:
\begin{equation}
\text{MB2}^{\Gamma^0}:\frac{\pi}{2}\tanh\left(a_2 \tilde\Gamma^0_{\mu\nu,\mu\nu}\right)
\end{equation}

Based on the form of the disconnected two-particle RDM, $\Gamma^0$, it is natural to consider a variant of the MB2 model in which its two constituent terms are treated as independent two-body descriptors and trained separately within the QML model:
\begin{equation}
\text{MB2}:\frac{\pi}{2}
\tanh\!\left(a_2 \tilde{P}_{\mu\mu}\tilde{P}_{\nu\nu}+b_2 \tilde{P}_{\mu\nu}\tilde{P}_{\mu\nu}\right)
\end{equation}
In the proposed MB2 model, we use only the disconnected two-particle reduced density matrix, which can be constructed directly from the one-particle density matrix. 

If the \emph{fully connected} two-particle RDM, $\Gamma_{\mu\mu',\nu\nu'}$, is available from a post-Hartree--Fock calculation, a potentially more accurate variant of the two-body layer can be constructed:
\begin{equation}
\text{MB2}^\Gamma:\frac{\pi}{2}
\tanh\!\left(a_2 \tilde\Gamma_{\mu\nu,\mu\nu}\right)
\end{equation}
Unlike MB2 or MB2$^{\Gamma^0}$, MB2$^\Gamma$ incorporates explicit electron-correlation information through the connected contribution to the full two-particle RDM. In this work, we construct the MB2$^{\Gamma}$ two-body encoding layer using the MP2 two-particle RDM.

These three two-body models provide a systematic hierarchy for incorporating electronic information into the two-qubit encoding layer. The MB2$^{\Gamma^0}$ model uses the disconnected two-particle RDM constructed from the one-particle density matrix and therefore captures pairwise information at the mean-field level. The second variant, MB2, introduces greater learning flexibility by treating the two constituent terms in $\Gamma^0$ independently, while retaining the two-qubit encoding. Finally, MB2$^{\Gamma}$ incorporates the correlated two-particle RDM obtained from MP2, introducing explicit electron-correlation effects into the two-body layer. These models allow us to assess how progressively richer one- and two-particle information affects the performance of the quantum encoding.

\begin{figure}[ht!]
\centering
\includegraphics[width=\linewidth]{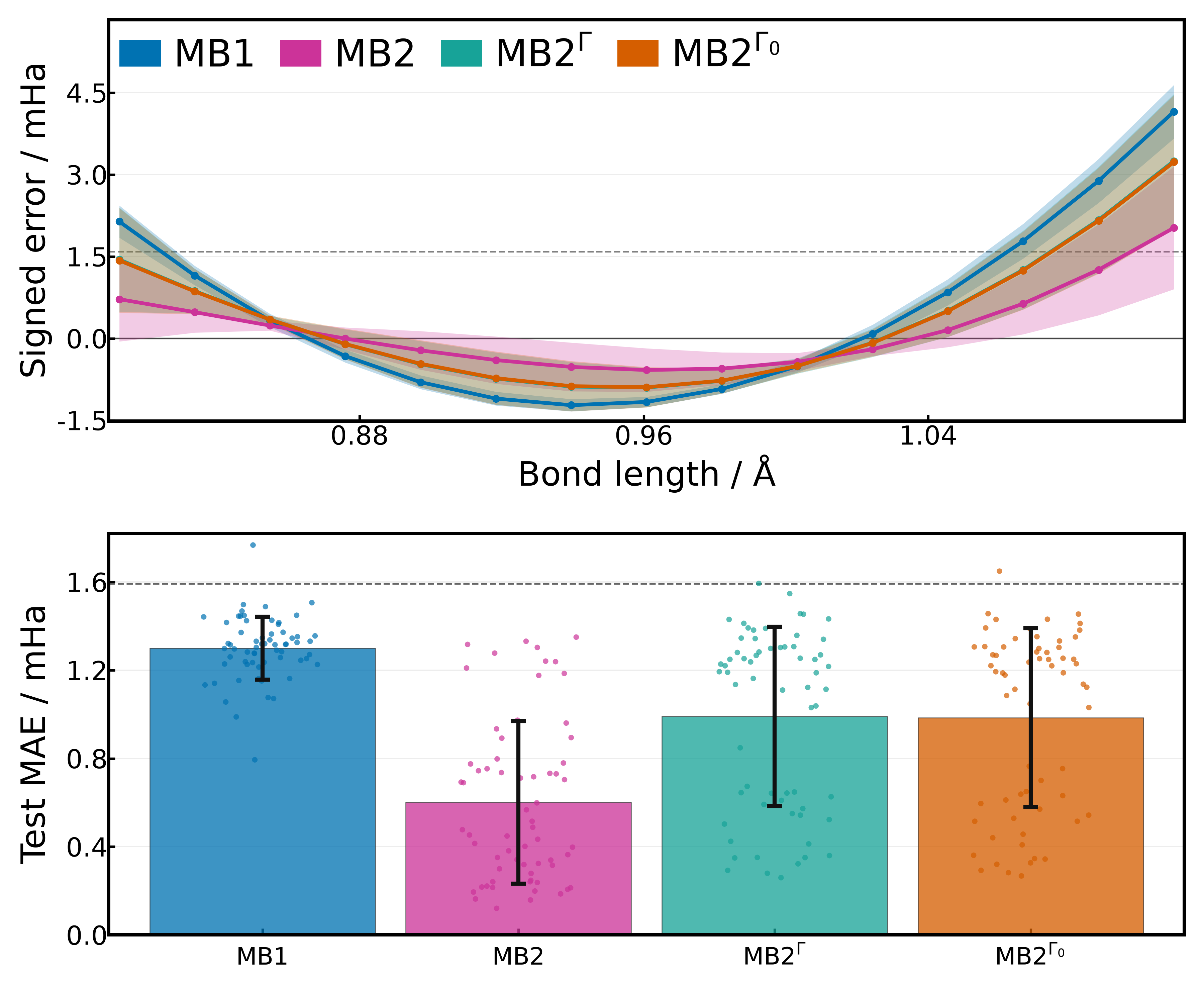}
\captionsetup{width=\linewidth}
\caption{Mean signed errors (top panel) and mean absolute errors (bottom panel) of the MB1, MB2,  MB2$^\Gamma$, and MB2$^{\Gamma^0}$ models for the HF molecule. Dashed horizontal lines mark the chemical-accuracy threshold of $\pm 1.6$ mHartree. The shaded area and error bar indicate standard error.}
\label{fig:MB2}
\end{figure}

\Cref{fig:MB2} shows that all three MB2 models show a significant improvement over the MB1 model for the HF molecule. Although only the diagonal elements of the two-particle RDM are considered, the MB2 encoding scheme introduces  two-electron information that is absent from the one-particle descriptors. This result indicates that even a reduced representation of the  two-particle density captures information relevant to the correlation energy that cannot be recovered from the one-body information alone.

\Cref{fig:MB2} shows that MB2$^\Gamma$ and MB2$^{\Gamma^0}$ achieve comparable accuracy and exhibit similar error distributions. This result is particularly noteworthy because MB2$^{\Gamma^0}$ relies only on the disconnected two-particle density, which can be constructed directly from the one-particle density matrix at substantially lower computational cost than the full two-particle RDM required by MB2$^\Gamma$. More importantly, it demonstrates that two-body electronic descriptors derived solely from one-particle density information can nevertheless encode sufficient information for the QML model to learn a significant portion of the electron-correlation effects.

Among the three MB2 models considered here, the two-parameter MB2 variant exhibits a noticeable improvement over both MB2$^\Gamma$ and MB2$^{\Gamma^0}$, yielding lower errors and a narrower error distribution. This improvement suggests that treating the two constituent terms of the disconnected two-particle RDM as independent descriptors provides additional flexibility for the QML model to learn their distinct contributions to electron correlation.

\subsubsection{Three-body and four-body layers}

The higher-order encoding layers involve the three- and four-particle RDMs, $\Gamma_{\mu\nu\lambda,\mu'\nu'\lambda'}$ and $\Gamma_{\mu\nu\lambda\kappa,\mu'\nu'\lambda'\kappa'}$, respectively. Computing the fully connected three- and four-particle RDMs is prohibitively expensive for all but very small systems. The analysis of the two-body layer, however, suggests that disconnected many-body RDMs constructed from lower-order density information may already provide sufficiently rich descriptors for QML. We therefore use disconnected higher-order density matrices to define the three- and four-body encoding layers, $\Gamma^0_{\mu\nu\lambda,\mu'\nu'\lambda'}$ in \cref{eq:gamma03} and $\Gamma^0_{\mu\nu\lambda\kappa,\mu'\nu'\lambda'\kappa'}$ in \cref{eq:gamma04}, avoiding the cost of explicitly evaluating the corresponding fully connected RDMs while retaining information about higher-order electronic couplings.

As in the MB2 layer, we restrict the three- and four-body encodings to the diagonal elements of the corresponding RDMs, enabling a direct mapping of the $n$-body descriptors onto $n$-qubit gates, leading to the MB3$^{\Gamma^0}$ and the MB4$^{\Gamma^0}$ models:
\begin{align}
\text{MB3}^{\Gamma^0}: &\frac{\pi}{2}\tanh\left(a_3 \tilde\Gamma^0_{\mu\nu\lambda,\mu\nu\lambda}\right)\label{eq:MB3gamma0}\\
\text{MB4}^{\Gamma^0}: &\frac{\pi}{2}\tanh\left(a_4 \tilde\Gamma^0_{\mu\nu\lambda\kappa,\mu\nu\lambda\kappa}\right)\label{eq:MB4gamma0}
\end{align}

As demonstrated for the MB2 model, the flexibility of QML training in the MB3$^{\Gamma^0}$ and MB4$^{\Gamma^0}$ models can be further enhanced by introducing independent trainable parameters for the constituent terms of the disconnected three- and four-particle RDMs, $\Gamma^0_{\mu\nu\lambda,\mu'\nu'\lambda'}$ and $\Gamma^0_{\mu\nu\lambda\kappa,\mu'\nu'\lambda'\kappa'}$, which contain three and five terms, respectively (see the full expressions in \cref{eq:gamma03} and \cref{eq:gamma04}):
\begin{align}
\text{MB3}: \frac{\pi}{2}\tanh&(a_3 \tilde{P}_{\mu\mu}\tilde{P}_{\nu\nu}\tilde{P}_{\lambda\lambda}\notag\\
&+b_3(\tilde{P}_{\mu\mu}\tilde{P}_{\nu\lambda}\tilde{P}_{\lambda\nu}+...)\notag\\&+c_3(\tilde{P}_{\mu\nu}\tilde{P}_{\nu\lambda}\tilde{P}_{\kappa\mu}+...))\label{eq:MB3gamma0}\\
\text{MB4}: \frac{\pi}{2}\tanh&(a_4 \tilde{P}_{\mu\mu}\tilde{P}_{\nu\nu}\tilde{P}_{\lambda\lambda}\tilde{P}_{\kappa\kappa}\notag\\
&+b_4(\tilde{P}_{\kappa\kappa}\tilde{P}_{\lambda\lambda}\tilde{P}_{\mu\nu}\tilde{P}_{\nu\mu}+...)\notag\\
&+c_4(\tilde{P}_{\kappa\lambda}\tilde{P}_{\lambda\kappa}\tilde{P}_{\mu\nu}\tilde{P}_{\nu\mu}+...)\notag\\
&+d_4(\tilde{P}_{\mu\nu}\tilde{P}_{\nu\lambda}\tilde{P}_{\nu\lambda}\tilde{P}_{\kappa\kappa}+...)\notag\\
&+e_4(\tilde{P}_{\mu\nu}\tilde{P}_{\nu\lambda}\tilde{P}_{\nu\kappa}\tilde{P}_{\kappa\mu}+...))\label{eq:MB4gamma0}
\end{align}

\begin{figure}[ht!]
\centering\includegraphics[width=\linewidth]{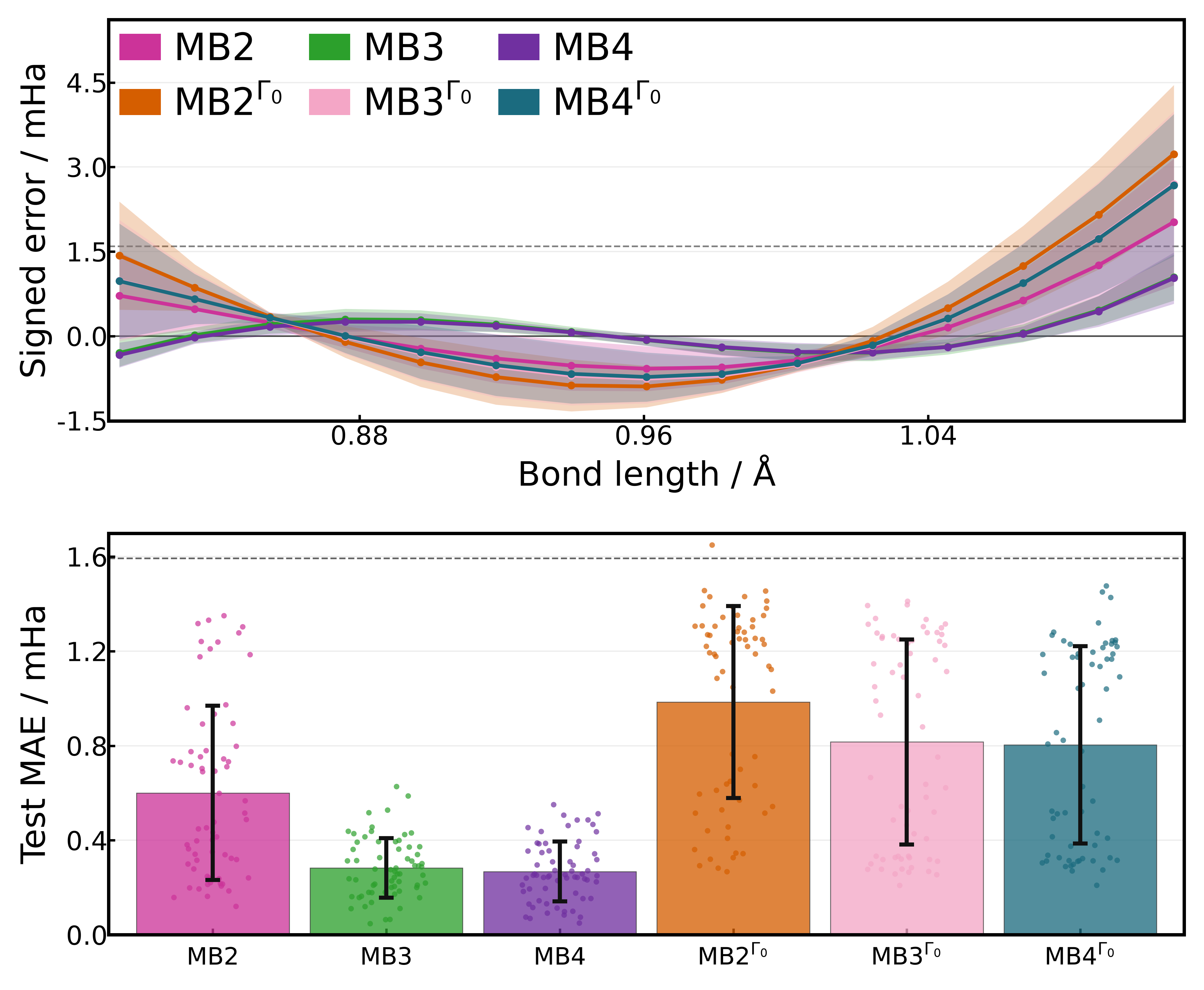}
\captionsetup{width=\linewidth}
\caption{Mean absolute errors of the MB3 and MB4 models for the HF molecule, compared with MB2. Dashed horizontal lines mark the chemical-accuracy threshold of $\pm 1.6$ mHartree. The shaded area and error bar indicate standard error.}
\label{fig:MB4}
\end{figure}

\Cref{fig:MB4} shows that all three- and four-body layers further reduce the QML prediction errors relative to the MB1 and MB2 models, although the incremental improvements are considerably smaller than those achieved by the lower-order encoding layers. These improvements suggest that higher-order many-body descriptors present the electronic structure information in a form not available from the one- and two-body descriptors, enabling the QML model to achieve a more accurate representation of electron-correlation effects. We again find that the more flexible MB3 and MB4 models outperform their corresponding single-parameter variants, MB3$^{\Gamma^0}$ and MB4$^{\Gamma^0}$, respectively. Not only does the MB4 model achieve the highest accuracy, with prediction errors well below the chemical-accuracy threshold across the entire range of bond lengths considered, but it also yields the narrowest error distribution.

\subsection{Transferability}

By comparing the extrapolation performance across different encoding orders, we further investigate the role of higher-body quantum gates in improving the robustness and transferability of quantum machine-learning predictions (see \cref{sec:transferability} for computational details). Specifically, this analysis examines whether the underlying many-body encodings capture transferable electronic-structure information that enables reliable predictions beyond the training domain, rather than merely interpolating correlation energies within the sampled bond-length range.

\begin{figure}[ht!]
\centering
\includegraphics[width=\linewidth]{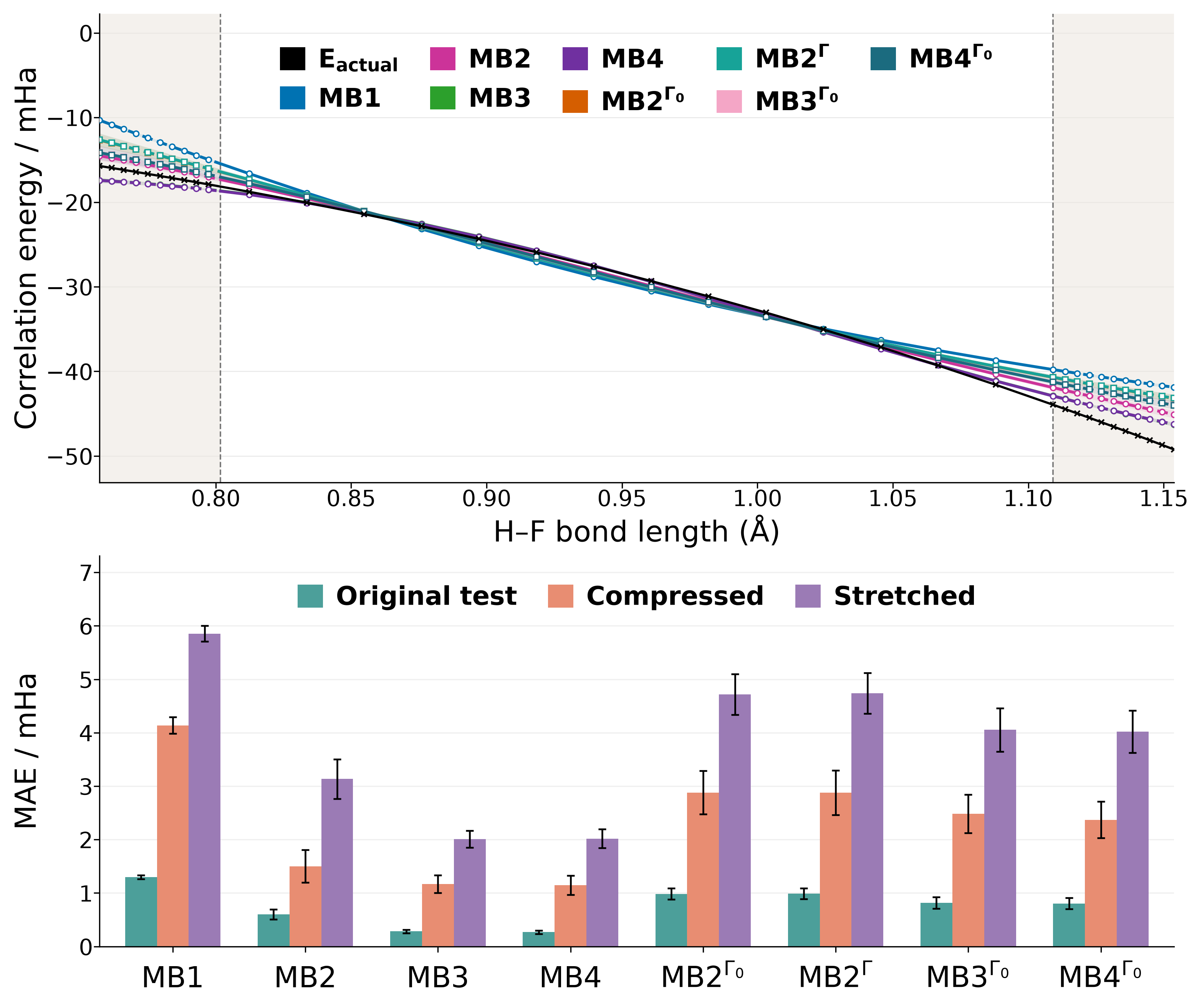}
\captionsetup{width=\linewidth}
\caption{Transferability Study: the testing range (shaded) is chosen to encompass the nuclear displacements associated with the second vibrationally excited state ($v=2$) on the Hartree--Fock potential energy surface. Ten testing geometries are selected on each side of the training range, corresponding to bond compression and elongation, respectively. {\bf{Top Panel}:} Correlation energy as a function of the HF bond length. {\bf{Bottom Panel}:} Mean absolute errors of the QML models for the HF molecule. The shaded area and error bar indicate standard error.
}
\label{fig:Transferability}
\end{figure}

As shown in \cref{fig:Transferability}, although prediction errors increase for molecular geometries outside the training region, the QML models largely retain their predictive accuracy. The prediction errors are larger in the bond-stretching region than in the bond-compression region. This behavior is expected, as electron-correlation effects become increasingly pronounced upon bond elongation, leading to larger correlation energies and a more challenging extrapolation regime for the QML models.

More importantly, the systematic, order-by-order improvement observed within the training domain persists in the extrapolation regime, with the MB4 model consistently exhibiting the best performance. These results suggest that higher-body quantum encodings not only enhance predictive accuracy but also improve the ability of QML models to generalize to previously unseen molecular geometries. The persistence of this hierarchical improvement further indicates that incorporating higher-order electronic-structure information into quantum circuit architectures provides a more transferable representation of electron correlation.

\section{Discussion}

The many-body encoding scheme exhibits systematic convergence as progressively higher-rank electronic information is incorporated. This hierarchy also maps naturally onto the structure of quantum circuits, where gates are classified by the number of qubits on which they act. An $n$-body electronic descriptor can therefore be associated directly with an $n$-qubit gate acting on the corresponding qubit registers.

In the many-body encoding scheme, molecular structure is not learned explicitly, as in the fingerprint model. Instead, structural information is encoded implicitly through the electronic density, since the electronic structure depends on the underlying nuclear geometry. Consequently, the hierarchical many-body representation can systematically improve predictive accuracy across different regions of a bond-stretch coordinate as progressively richer electronic information is incorporated.

An interesting finding from the many-body encoding analysis is that QML models trained on disconnected $n$-particle RDMs can nevertheless accurately predict electron-correlation corrections. From the perspective of correlated electronic structure theory, disconnected $n$-particle RDMs do not explicitly capture genuine $n$-particle correlations, as they are constructed entirely from lower-order density information. However, their role in QML is fundamentally different. Although disconnected RDMs do not explicitly encode electron correlation, they preserve important structural information inherited from the one-particle density, including orbital occupations, orbital coherences, and their higher-order products. As molecular geometry changes, these quantities vary systematically, providing physically meaningful descriptors that reflect changes in the underlying electronic structure. A QML model can therefore learn the relationship between these descriptors and electron-correlation corrections, even when the descriptors themselves do not explicitly contain the corresponding correlation contributions.

The number of gates grows as $\binom{N}{n}$ at the $n$-body level, which can become resource demanding for large systems. As demonstrated by MB2, however, including only selected diagonal elements of the two-particle density can already substantially improve the QML prediction. For molecular systems dominated by dynamic correlation, this observation suggests a clear path toward incorporating additional correlation information without retaining the full high-rank tensor.

\section{Conclusion and Perspective}

We have shown that the body order of electronic-structure descriptors can be matched directly to the locality of quantum gates. One-body populations and orbital energies set single-qubit rotations, and the diagonal elements of the disconnected two-, three-, and four-particle reduced density matrices set two-, three-, and four-qubit interaction strengths. All descriptors come from a single Hartree--Fock calculation, and no fermion-to-qubit mapping or Pauli decomposition is needed. Across bond-stretching scans of five molecules, the resulting hierarchy converges systematically toward the full configuration interaction correlation energy, with the largest gains coming from the one- and two-body layers and smaller but consistent gains from the three- and four-body layers.

The strong performance of QML models using descriptors constructed from disconnected many-particle densities suggests that explicitly correlated inputs are not necessary for learning electron-correlation effects. Instead, the key requirement is that the descriptors capture the underlying many-body structure relevant to the target physics. Because disconnected densities can be constructed directly from the mean-field one-particle density at negligible additional computational cost, this finding provides a practical and computationally efficient route to incorporating higher-body encoding layers in QML models.

The encoding also fixes the number of trainable parameters independent of molecular size. Shared mapping functions turn descriptors into angles, and two shared rotation angles form the readout. The models place the burden of expressivity on the encoding rather than on the ansatz, and keeps the circuit shallow enough.

Finally, the transferability analysis demonstrates that the hierarchical improvement achieved through higher-body quantum encodings persists beyond the training domain, with the MB4 model consistently exhibiting the best extrapolation performance. Although prediction errors increase at more strongly stretched bond geometries, where electron-correlation effects become more pronounced, the QML models largely retain their predictive accuracy.

Several questions remain open and define the next stage of this work. The present benchmarks use one-dimensional scans with per-molecule training, a setting in which smooth interpolation is possible for many model classes. A direct comparison against classical models trained on the same reduced-density-matrix descriptors is needed to separate the contribution of the descriptors from that of the quantum encoding, and we are pursuing this alongside multidimensional potential energy surfaces where the interpolation problem is harder. 

The size-independent parametrization invites transfer learning: a model trained on one set of molecules and applied to another, or trained at a few geometries and applied across a full surface. Establishing whether the shared parameters capture transferable physics is, in our view, the most important test of the scheme. 

The gate count grows combinatorially with the number of orbitals, and a locality-based screening of orbital pairs, triples, and quadruples, following the known nearsightedness of electron correlation, will be required for larger systems; the two-body results suggest that selected diagonal elements suffice. 

Finally, all results here are from noiseless simulation, and the robustness of the hierarchy under realistic noise models and on hardware remains to be characterized.

More broadly, the principle is not specific to correlation energy. Any target with a many-body hierarchy, from excited-state energies to response properties, admits the same construction: identify the physical quantities of each body order, assign them to gates of matching locality, and let the circuit architecture carry the structure of the data. Treating the encoding as a physical statement rather than a preprocessing step is, we believe, a general route to compact and interpretable quantum machine-learning models for chemistry.

\section{Methods}
\label{sec:methods}

The QML task explored in this work is the prediction of the electron correlation energy. Electron correlation represents the many-body contribution to the electronic energy that is missing from a mean-field description. Because correlation effects arise from collective interactions among electrons and vary nontrivially with molecular geometry, they provide a natural test bed for assessing whether a quantum encoding can capture physically meaningful many-body information. We therefore use the correlation energy as the learning target and examine how progressively richer one-body and many-body descriptors improve the expressiveness of the QML model.

\subsection{Many-Body Descriptor Construction}
\label{sec:mb_descriptor}

Throughout this section, $i,j,k,l$ denote molecular orbitals (MOs), whereas
$\mu,\nu,\lambda,\kappa$ denote atomic orbitals (AOs).

\subsubsection{Hartree--Fock one-particle quantities}

For each molecular geometry, a Hartree--Fock calculation is first performed. The corresponding one-particle density matrix in the AO basis is
\begin{equation}
  P_{\mu\nu}
  =
  2\sum_i^{\mathrm{occ}}
  C_{\mu i}C_{\nu i}^{*},
  \label{eq:rhf_density}
\end{equation}
where the sum runs over the occupied spatial MOs and $C$'s are MO coefficients.

\subsubsection{Connected and disconnected two-particle reduced density matrix}

The fully connected two-particle reduced density matrix (RDM) is obtained from a correlated electronic-structure calculation. In this work, we use the MP2\cite{Plesset34_618,Seeger76_1} two-particle RDM in the canonical molecular-orbital basis, $\Gamma_{ijkl}$, which is subsequently transformed to the atomic-orbital basis:
\begin{equation}
\Gamma_{\mu\nu,\mu'\nu'}
  =
  \sum_{ijkl}
  C_{\mu i}
  C_{\nu j}
  C_{\mu' k}
  C_{\nu' l}
  \Gamma_{ijkl}.
  \label{eq:gamma2}
\end{equation}

The full two-particle RDM contains both a disconnected contribution, which can be constructed from products of the one-particle density matrix, and a connected contribution that captures genuine two-particle correlation.

The disconnected two-particle RDM can  be easily constructed directly from the Hartree--Fock one-particle density matrix:\cite{Mukherjee99_2800}
\begin{equation}
  \Gamma^{0}_{\mu\nu,\mu'\nu'}
  =
  P_{\mu\mu'}P_{\nu\nu'}
  -
  \frac{1}{2}
  P_{\mu\nu'}P_{\nu\mu'}.
  \label{eq:gamma02}
\end{equation}

\subsubsection{Three-particle disconnected reduced density matrix}\label{sec:three-body}

For a restricted Hartree--Fock reference, the spin-summed disconnected three-particle reduced density matrix is
\begin{align}
&\Gamma^{0}_{\mu\nu\kappa,\mu'\nu'\kappa'} = P_{\mu\mu'}P_{\nu\nu'}P_{\kappa\kappa'} \notag\\ \nonumber
  &\quad - \frac{1}{2}[P_{\mu\mu'}P_{\nu\kappa'}P_{\kappa\nu'} + P_{\nu\nu'}P_{\mu\kappa'}P_{\kappa\mu'} + P_{\kappa\kappa'}P_{\mu\nu'}P_{\nu\mu'}] \\ 
  &\quad + \frac{1}{4}[P_{\mu\nu'}P_{\nu\kappa'}P_{\kappa\mu'} + P_{\mu\kappa'}P_{\nu\mu'}P_{\kappa\nu'}]
  \label{eq:gamma03}
\end{align}

\subsubsection{Four-particle disconnected reduced density matrix}\label{sec:four-body}

For a restricted Hartree--Fock reference, the spin-summed disconnected four-particle reduced density matrix is
\begin{align}
&\Gamma^0_{\mu\nu\kappa\lambda,\mu'\nu'\kappa'\lambda'} = P_{\mu\mu'}P_{\nu\nu'}P_{\kappa\kappa'}P_{\lambda\lambda'}\notag\\
& \quad -
\frac{1}{2}\Big[P_{\kappa\kappa'}P_{\lambda\lambda'}P_{\mu\nu'}P_{\nu\mu'}
+P_{\nu\nu'}P_{\lambda\lambda'}P_{\mu\kappa'}P_{\kappa\mu'}\notag\\
&\quad +P_{\nu\nu'}P_{\kappa\kappa'}P_{\mu\lambda'}P_{\lambda\mu'}
+P_{\mu\mu'}P_{\lambda\lambda'}P_{\nu\kappa'}P_{\kappa\nu'}
\notag\\
&\quad + P_{\mu\mu'}P_{\kappa\kappa'}P_{\nu\lambda'}P_{\lambda\nu'}+P_{\mu\mu'}P_{\nu\nu'}P_{\kappa\lambda'}P_{\lambda\kappa'}\Big]
\notag\\
& \quad +
\frac{1}{4}\Big[P_{\mu\nu'}P_{\nu\mu'}P_{\kappa\lambda'}P_{\lambda\kappa'}
+P_{\mu\kappa'}P_{\kappa\mu'}P_{\nu\lambda'}P_{\lambda\nu'} \notag\\
&\quad + P_{\mu\lambda'}P_{\lambda\mu'}P_{\nu\kappa'}P_{\kappa\nu'}\Big]
\notag\\
& \quad +  \frac{1}{4}\Big[P_{\lambda\lambda'}\big(P_{\mu\nu'}P_{\nu\kappa'}P_{\kappa\mu'}+P_{\mu\kappa'}P_{\kappa\nu'}P_{\nu\mu'}\big)\notag\\
&\quad + P_{\kappa\kappa'}\big(P_{\mu\nu'}P_{\nu\lambda'}P_{\lambda\mu'}+P_{\mu\lambda'}P_{\lambda\nu'}P_{\nu\mu'}\big)\notag\\
&\quad + P_{\nu\nu'}\big(P_{\mu\kappa'}P_{\kappa\lambda'}P_{\lambda\mu'}+P_{\mu\lambda'}P_{\lambda\kappa'}P_{\kappa\mu'}\big)
\notag\\
&\quad + P_{\mu\mu'}\big(P_{\nu\kappa'}P_{\kappa\lambda'}P_{\lambda\nu'}+P_{\nu\lambda'}P_{\lambda\kappa'}P_{\kappa\nu'}\big)\Big]
\notag\\ 
&\quad - 
\frac{1}{8}\Big[P_{\mu\nu'}P_{\nu\kappa'}P_{\kappa\lambda'}P_{\lambda\mu'}   + P_{\mu\nu'}P_{\nu\lambda'}P_{\lambda\kappa'}P_{\kappa\mu'}
\notag\\
&\quad + P_{\mu\kappa'}P_{\kappa\nu'}P_{\nu\lambda'}P_{\lambda\mu'}+ P_{\mu\kappa'}P_{\kappa\lambda'}P_{\lambda\nu'}P_{\nu\mu'}
\notag\\ 
&\quad +P_{\mu\lambda'}P_{\lambda\nu'}P_{\nu\kappa'}P_{\kappa\mu'}+ P_{\mu\lambda'}P_{\lambda\kappa'}P_{\kappa\nu'}P_{\nu\mu'}\Big]\label{eq:gamma04}
\end{align}

\subsubsection{L\"owdin transformation}

The AO basis is nonorthogonal. Before constructing the final
descriptors, the one- and two-particle quantities are transformed to
a symmetrically orthogonalized L\"owdin basis.\cite{Lowdin50_365,Li03_7651} 
The one-particle density matrices is transformed according to
\begin{align}
  \mathbf{P}^{L}
  &=
  \mathbf{S}^{1/2}
  \mathbf{P}
  \mathbf{S}^{1/2},
  \label{eq:density_lowdin}
\end{align}
where the superscript $L$ denotes a quantity represented in the
L\"owdin-orthogonalized  AO basis.

Each index of the rank-four two-particle RDM is transformed with
$\mathbf{S}^{1/2}$:
{\footnotesize
\begin{align}
  &\Gamma_{\mu\nu\lambda\kappa}^L
  =\notag\\
  &\sum_{\mu'\nu'\lambda'\kappa'}
  (\mathbf{S}^{1/2})_{\mu\mu'}
  (\mathbf{S}^{1/2})_{\nu\nu'}
  (\mathbf{S}^{1/2})_{\lambda\lambda'}
  (\mathbf{S}^{1/2})_{\kappa\kappa'}
  \Gamma^{\mathrm{AO}}_{\mu'\nu'\lambda'\kappa'}.
  \label{eq:cumulant_lowdin}
\end{align}
}

In the following discussion, the superscript $L$ is omitted for simplicity. Unless otherwise noted, all many-body encoding descriptors are expressed in the orthonormal L\"owdin basis.

\subsection{Quantum Machine-Learning Many-body Encoding Scheme}\label{sec:many-body}

Because core electrons contribute substantially less to the correlation energy than valence electrons, chemically identified core AO functions are excluded from the descriptor space. For example, the $1s$ functions of second-row atoms are treated as core and omitted from the encoding. Owing to their strong spatial localization and energetic separation from the valence space, these compact core functions largely retain their atomic-orbital character after L\"owdin orthonormalization, making their identification in the orthonormal AO basis straightforward.

Each selected L\"owdin atomic orbital is assigned one qubit, and the circuit starts from $|0\rangle^{\otimes N}$ where $N$ is the number of orbitals. Four many-body encoding layers follow, one for each class of descriptor.

\subsubsection{Standardization}\label{sec:standardization}

The one-particle L\"owdin density matrix elements are first standardized/normalized according to
\begin{align}
\tilde{P}_{\mu\mu} &=
\frac{P_{\mu\mu}-\bar{P}_{\mu\mu}}{\sigma_{P_{\mu\mu}}},\\
\tilde{P}_{\mu\nu} &=
\frac{P_{\mu\nu}-\bar{P}_{\mu\nu}|_{\mu\ne\nu}}{\sigma_{P_{\mu\nu}|_{\mu\ne\nu}}},
\label{eq:standardize}
\end{align}
where $\bar{P}$ and $\sigma_P$  denote the mean and standard deviation of the corresponding descriptor over the training geometries for a given molecule.

Similarly, $n$-body RDMs are standardized according to:
\begin{align}
\tilde{\Gamma}^0_{\mu\nu,\mu\nu} &=
\frac{\Gamma^0_{\mu\nu,\mu\nu}-\bar{\Gamma}^0_{\mu\nu,\mu\nu}}{\sigma_{\Gamma^0_{\mu\nu,\mu\nu}}},\\
\tilde{\Gamma}_{\mu\nu,\mu\nu} &=
\frac{\Gamma_{\mu\nu,\mu\nu}-\bar{\Gamma}_{\mu\nu,\mu\nu}}{\sigma_{\Gamma_{\mu\nu,\mu\nu}}},\\
\tilde{\Gamma}^0_{\mu\nu\lambda,\mu\nu\lambda} &=\frac{\Gamma^0_{\mu\nu\lambda,\mu\nu\lambda}-\bar{\Gamma}^0_{\mu\nu\lambda,\mu\nu\lambda}}{\sigma_{\Gamma^0_{\mu\nu\lambda,\mu\nu\lambda}}},\\
\tilde{\Gamma}^0_{\mu\nu\lambda\kappa,\mu\nu\lambda\kappa} &=\frac{\Gamma^0_{\mu\nu\lambda\kappa,\mu\nu\lambda\kappa}-\bar{\Gamma}^0_{\mu\nu\lambda\kappa,\mu\nu\lambda\kappa}}{\sigma_{\Gamma^0_{\mu\nu\lambda\kappa,\mu\nu\lambda\kappa}}},
\end{align}
where $\bar{\Gamma^0}$ ($\bar{\Gamma}$) and $\sigma_{\Gamma^0}$ ($\sigma_\Gamma$) are the mean and standard deviation. 

Standardizing the density-matrix descriptors substantially improves the QML performance, although it does not introduce additional electronic-structure information. Rather, the transformation removes the nearly geometry-independent background in the orbital populations and rescales their variations to comparable magnitudes. The resulting descriptors therefore emphasize the redistribution of electronic density along the molecular geometry, which provides the relevant signal for learning changes in electron correlation. Standardization also improves the numerical conditioning of the quantum encoding: because the descriptors are centered near zero with approximately unit variance, they predominantly sample the high-sensitivity region of the nonlinear mapping from density descriptors to gate angles. This effect is particularly important when encoding parameters are shared across orbitals, as it allows a common trainable parameter to respond consistently to descriptors with otherwise different magnitudes and variances.

\subsubsection{One-body (MB1) layer}\label{sec:MB1}
The one-body layer acts independently on each qubit and uses the diagonal element of the standardized L\"owdin density matrix, $\tilde{P}_{\mu\mu}$. It tells how much electron density is associated with orbital $\mu$. 

For the qubit associated with orbital $\mu$, the rotation angle is defined as
\begin{equation}
\gamma_{\mu} =
\frac{\pi}{2}
\tanh\!\left(
a_{1}\tilde{P}_{\mu\mu}
\right),
\label{eq:gamma1}
\end{equation}
and is encoded through an $R_Y(\gamma_{\mu})$ rotation. The coefficient $a_1$ is a trainable parameter shared across all qubits.
The $\tanh$ transformation bounds the rotation angle to $(-\pi/2,\pi/2)$, preventing periodic wrapping of the rotation and preserving a one-to-one mapping between the descriptor and the encoded angle.

\subsubsection{Two-body (MB2) layer}\label{sec:MB2}
The two-body layer acts on pairs of qubits associated with orbitals $\mu$ and $\nu$. The diagonal elements of the connected (\cref{eq:gamma2}) or disconnected (\cref{eq:gamma02}) two-particle reduced density matrix  are used as the two-body descriptors. For each orbital pair, the descriptor is mapped to an interaction angle,
\begin{equation}
\gamma_{\mu\nu}=
\frac{\pi}{2}
\tanh\!\left(a_2 \tilde\Gamma_{\mu\nu,\mu\nu}\right),
\label{eq:MB2encode}
\end{equation}
or
\begin{equation}
\gamma_{\mu\nu}=
\frac{\pi}{2}
\tanh\!\left(a_2 \tilde\Gamma^0_{\mu\nu,\mu\nu}\right),
\label{eq:MB2encode}
\end{equation}
where $a_2$ is a trainable parameter shared across all orbital pairs. 

We also consider a variant of the two-body layer in which its two constituent terms in $\Gamma^0$ are treated as independent two-body descriptors and trained separately within the QML model:
\begin{equation}
\gamma_{\mu\nu}=\frac{\pi}{2}
\tanh\!\left(a_2 \tilde{P}_{\mu\mu}\tilde{P}_{\nu\nu}+b_2 \tilde{P}_{\mu\nu}\tilde{P}_{\mu\nu}\right)
\end{equation}

The corresponding two-qubit gate for all MB2 models is 
\begin{equation}
U_{\mu\nu}
=\exp\!\left(
-\frac{i}{2}\gamma_{\mu\nu} Z_{\mu}Z_{\nu}
\right).
\label{eq:two_body_gate}
\end{equation}
Thus, the strength of the $ZZ$ interaction between qubits $\mu$ and $\nu$ is determined directly by the electronic coupling encoded in the two-particle RDM. As in the one-body layer, the $\tanh$ function bounds the interaction angle within $(-\pi/2,\pi/2)$ and prevents periodic wrapping of the encoded descriptor. Because the same coefficients $a_2$ and $b_2$ are used for every pair, the two-body layer introduces only one to two additional trainable parameters independent of system size.

\subsubsection{Three-body (MB3) layer}\label{sec:MB3}

The three-body layer acts on triples of qubits associated with orbitals $\mu$, $\nu$, and $\lambda$. The corresponding three-body descriptor is derived from the diagonal elements of the disconnected three-particle RDM defined in \cref{eq:gamma03}. For each orbital triple, the diagonal element $\Gamma^0_{\mu\nu\lambda,\mu\nu\lambda}$ or its more flexible variant is mapped to an interaction angle,
\begin{align}
\gamma_{\mu\nu\lambda}
=
\frac{\pi}{2}
\tanh\!(&a_3 \tilde\Gamma^0_{\mu\nu\lambda,\mu\nu\lambda}),\\
\gamma_{\mu\nu\lambda}=\frac{\pi}{2}\tanh&(a_3 \tilde{P}_{\mu\mu}\tilde{P}_{\nu\nu}\tilde{P}_{\lambda\lambda}\notag\\
&+b_3(\tilde{P}_{\mu\mu}\tilde{P}_{\nu\lambda}\tilde{P}_{\lambda\nu}+...)\notag\\
&+c_3(\tilde{P}_{\mu\nu}\tilde{P}_{\nu\lambda}P_{\kappa\mu}+...))
\end{align}
where $a_3$, $b_3$, $c_3$ are trainable parameters shared across all orbital triples (see \cref{eq:gamma03} for the full expression). The corresponding three-qubit interaction gate is
\begin{equation}
U_{\mu\nu\lambda}
=
\exp\!\left(
-\frac{i}{2}
\gamma_{\mu\nu\lambda}
Z_{\mu}Z_{\nu}Z_{\lambda}
\right).
\label{eq:three_body_gate}
\end{equation}
In this way, each diagonal element of the disconnected three-particle RDM directly determines the strength of the corresponding three-qubit interaction, allowing higher-order electronic information to modulate the three-body entangling layer.

\subsubsection{Four-body (MB4) layer}\label{sec:MB3}

The four-body layer acts on quadruples of qubits associated with orbitals $\mu$, $\nu$, $\lambda$, and $\kappa$. The corresponding four-body descriptor is derived from the diagonal elements of the disconnected four-particle RDM defined in \cref{eq:gamma04}. For each orbital quadruple, the diagonal element $\Gamma^0_{\mu\nu\lambda\kappa,\mu\nu\lambda\kappa}$ or its more flexible variant is mapped to an interaction angle,
\begin{align}
\gamma_{\mu\nu\lambda\kappa}
=
\frac{\pi}{2}
\tanh\!(
&a_4 \tilde\Gamma^0_{\mu\nu\lambda\kappa,\mu\nu\lambda\kappa}),
\label{eq:gamma4}\\
\gamma_{\mu\nu\lambda\kappa}=\frac{\pi}{2}\tanh&(a_4 \tilde{P}_{\mu\mu}\tilde{P}_{\nu\nu}\tilde{P}_{\lambda\lambda}\tilde{P}_{\kappa\kappa}\notag\\
&+b_4(\tilde{P}_{\kappa\kappa}\tilde{P}_{\lambda\lambda}\tilde{P}_{\mu\nu}\tilde{P}_{\nu\mu}+...)\notag\\
&+c_4(\tilde{P}_{\kappa\lambda}\tilde{P}_{\lambda\kappa}\tilde{P}_{\mu\nu}\tilde{P}_{\nu\mu}+...)\notag\\
&+d_4(\tilde{P}_{\mu\nu}\tilde{P}_{\nu\lambda}\tilde{P}_{\nu\lambda}\tilde{P}_{\kappa\kappa}+...)\notag\\
&+e_4(\tilde{P}_{\mu\nu}\tilde{P}_{\nu\lambda}\tilde{P}_{\nu\kappa}\tilde{P}_{\kappa\mu}+...))
\end{align}
where $a_4$, $b_4$, $c_4$, $d_4$, and $e_4$ are trainable parameters shared across all orbital quadruples (see \cref{eq:gamma04} for the full expression). The corresponding four-qubit interaction gate is
\begin{equation}
U_{\mu\nu\lambda\kappa}
=
\exp\!\left(
-\frac{i}{2}
\gamma_{\mu\nu\lambda\kappa}
Z_{\mu}Z_{\nu}Z_{\lambda}Z_{\kappa}
\right).
\label{eq:four_body_gate}
\end{equation}

\subsubsection{Trainable ansatz and readout}

The two-, three-, and four-index information encoded by the $ZZ$, $ZZZ$, and $ZZZZ$ gates is not directly accessible through $Z$-basis measurements. Because these interaction gates commute with every $Z_\mu$, they modify relative phases without changing the qubit populations, leaving $\langle Z_\mu\rangle$ unchanged immediately after encoding. To convert this phase information into measurable population differences, the encoding is followed by two trainable layers. Each layer applies an $R_Y(\theta_l)$ rotation with a single angle $\theta_l$ shared across all qubits, followed by a ring of CNOT gates,
\begin{equation}
U_{\mathrm{train}}^{(l)}
= U_{\mathrm{CNOT}}^{\mathrm{ring}}
\prod_{\mu\in\mathcal{A}} R_Y^{(\mu)}(\theta_l),
\qquad l=1,2.
\label{eq:trainable_layer}
\end{equation}

Because $R_Y$ does not commute with $Z$, these rotations mix computational-basis components carrying different descriptor-dependent phases and convert the encoded phase information into population differences. The subsequent CNOT ring propagates this information among the qubits and introduces additional interqubit correlations. Repeating the block twice provides a shallow variational circuit with only two trainable parameters, $\theta_1$ and $\theta_2$, independent of the number of orbitals. The trainable block therefore does not introduce additional physical descriptors; instead, it learns how to transform the one-, two-, three-, and four-body information already encoded in the quantum state into observables accessible through the final $Z$-basis measurement.

Expressivity is thus placed deliberately in the encoding rather than in the ansatz. This keeps the circuit shallow, at the cost of limiting what the trainable block can correct if the encoding is incomplete. The ansatz and readout are the same for all MB$n$ models, so differences between them come from the encoding layers rather than from a more expressive ansatz.

After the trainable block, each qubit is measured in the computational
basis to give $\langle Z_\mu \rangle$. The predicted correlation energy is
obtained from these expectation values as
\begin{equation}
  E_{\mathrm{pred}} = \beta + \sum_{\mu=1}^{K} w_\mu \bar{z}_\mu
  + \sum_{k=1}^{3} u_k M_k,
  \label{eq:readout}
\end{equation}
where $N$ is the number of qubits, $\bar{z}$ collects the
$\langle Z_\mu \rangle$ padded with zeros to length $K = \max(8, N)$, $M_k = N^{-1} \sum_{\mu} \langle Z_\mu \rangle^{k}$ is the $k$th moment of the expectation values, and $\beta$, $w_\mu$, and $u_k$ are trainable weights. The weights are needed because $\langle Z_\mu \rangle \in [-1,1]$, whereas the correlation energy is tens to hundreds of mHartree. Models are trained by minimizing the mean squared error between $E_{\mathrm{pred}}$ and $E_{\mathrm{corr}}$ over the training geometries.

\subsection{Computational Details}

\subsubsection{Electronic structure calculations}\label{sec:abintio}

All calculations employ the STO-3G minimal basis set
\cite{Pople69_2657}. Its compact size makes full configuration interaction (FCI) calculations tractable for every molecule considered, providing correlation-energy targets that are exact within the chosen basis and frozen-core approximation. The basis also determines the qubit-register size, since the encoding assigns one qubit to each valence atomic orbital.

The two-particle information used by the encoding is obtained from second-order M{\o}ller--Plesset perturbation theory (MP2) based on the Hartree--Fock reference \cite{Plesset34_618}.

\subsubsection{Molecular systems}\label{sec:molecules}

Five molecules (HF, H$_2$O, NH$_3$, N$_2$, and CO) are studied through bond-stretching scans. 
Symmetry-equivalent bonds are stretched simultaneously to preserve the molecular point-group symmetry. Each scan comprises 30 evenly spaced geometries near the equilibrium structure.

\begin{figure}[ht!]
\centering
\includegraphics[width=\linewidth]{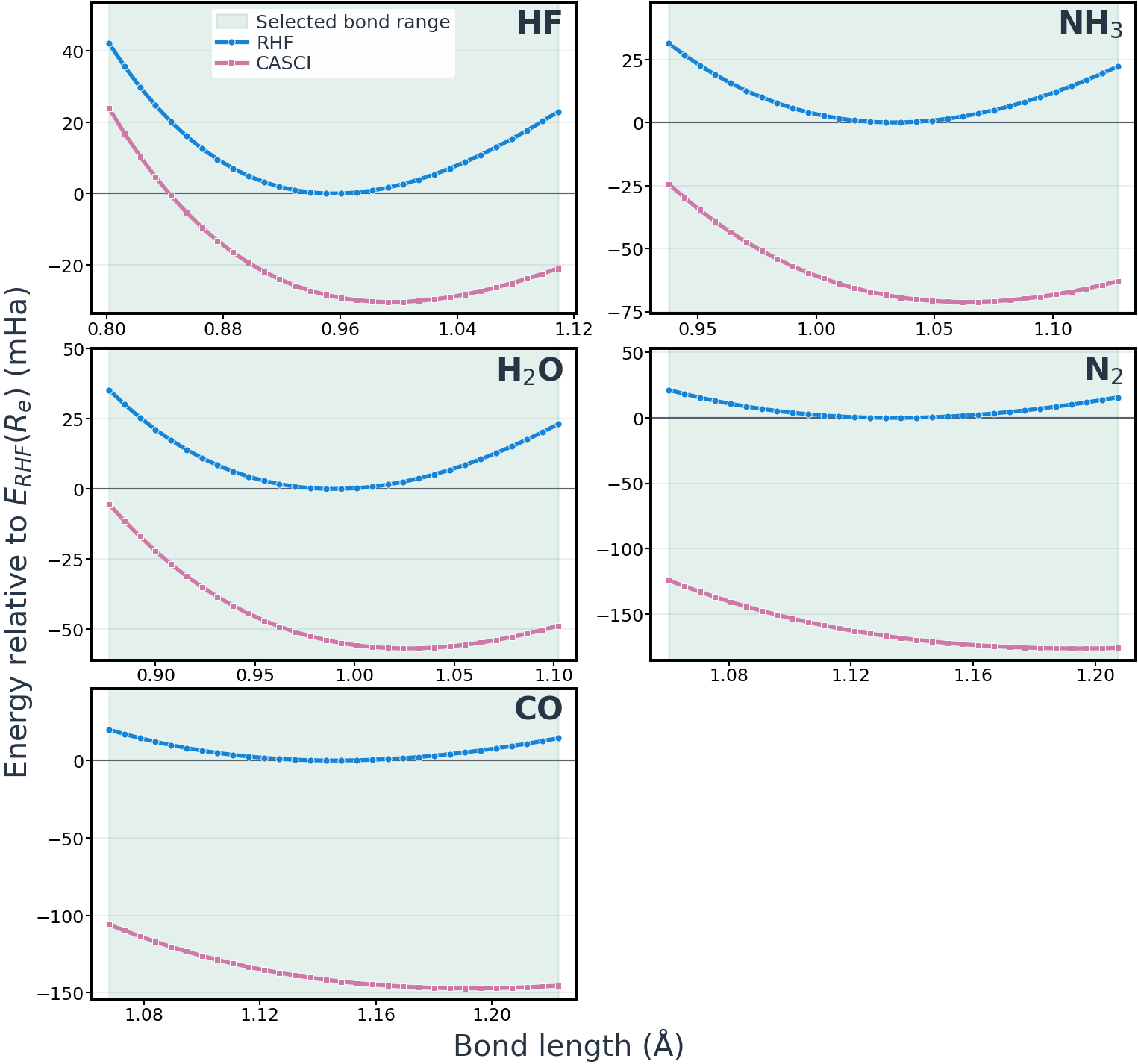}
\captionsetup{width=\linewidth}
\caption{Hartree--Fock and CI potential energy curves for five molecules along the bond-stretching coordinates used to train the QML models. Symmetry-equivalent bonds are varied simultaneously. The energy difference between the FCI and Hartree--Fock curves defines the correlation-energy target.}
\label{fig:PES}
\end{figure}

\Cref{fig:PES} shows the Hartree--Fock and CI potential energy curves over the structural range used to train the QML models. The structural range is chosen to span the nuclear displacements associated with the first vibrationally excited state ($v=1$) on the Hartree--Fock potential energy surface, extending beyond those associated with zero-point motion.

\Cref{fig:PES} also shows significant variation in correlation energy, bonding characteristics, and molecular symmetry (C$_\infty$, C$_2$, C$_3$) across the five molecules, providing a diverse dataset for evaluating the QML encoding schemes and models. Molecules with multiple bonds, such as CO and N$_2$, exhibit larger correlation corrections relative to the Hartree--Fock reference than the other single-bonded molecules considered here. For every molecule, the magnitude of the correlation energy is smaller at the compressed end of the scan than at the stretched end. 

\subsubsection{QML setup}

QML models are trained separately for each molecule with the Adam optimizer,\cite{Ba15_1412.6980,Li25_026110} at a learning rate of 0.02 for 1,000 epochs. The test MAE is computed for each seed and averaged over 64 seeds. Reported values are means over the 64 seeds.

For each molecule, the 30 geometries are partitioned into 15 training and 15 test geometries using an alternating split. All circuits are simulated as exact statevectors in Pennylane,
without shot noise or hardware noise. Noise-free simulation isolates the effect of the encoding; robustness under noise is left to future work.

\Cref{tab:gates} summarizes the number of qubits and encoding gates in each layer for all molecules, while \Cref{tab:params} reports the number of trainable parameters in each model.

\begin{table}
  \captionsetup{width=\linewidth}
  \caption{Number of qubits, $N$, and encoding gates in each layer. Each qubit represents one L\"owdin atomic orbital. The two-, three-, and four-body layers act on all distinct pairs, triples, and quadruples of qubits, respectively. The MB$n$ model includes all encoding layers up to order $n$.}
  \label{tab:gates}
  \centering
  \begin{tabular}{lccccc}
    \toprule
    Molecule & $N$ & $R_Y$ & $ZZ$ & $ZZZ$ & $ZZZZ$ \\
    \midrule
    HF         & 5  & 5  & 10 & 10  & 5   \\
    H$_2$O     & 6  & 6  & 15 & 20  & 15  \\
    NH$_3$     & 7  & 7  & 21 & 35  & 35  \\
    N$_2$      & 8  & 8  & 28 & 56  & 70  \\
    CO         & 8  & 8  & 28 & 56  & 70  \\
    \bottomrule
  \end{tabular}
\end{table}

\begin{table}
  \captionsetup{width=\linewidth}
  \caption{Number of trainable parameters in each model. The encoding column reports the trainable coefficients associated with the encoding layers. Each model also includes two ansatz parameters, $\theta_1$ and $\theta_2$, together with $N+4$ readout parameters, $\beta$, $w_\mu$, and $u_k$. The final column gives the corresponding parameter counts for CO, which is the largest molecular system considered here, for which $N=8$.}
  \label{tab:params}
  \centering
  \begin{tabular}{lccc}
    \toprule
    Model & Encoding & Total & CO \\
    \midrule
    MB1              & 1  & $N+7$  & 15 \\
    MB2              & 3  & $N+9$ & 17 \\
    MB2$^{\Gamma^0}$ & 2  & $N+8$ & 16 \\
    MB2$^{\Gamma}$   & 2  & $N+8$ & 16 \\
    MB3              & 6  & $N+12$ & 20 \\
    MB3$^{\Gamma^0}$ & 3  & $N+9$ & 17 \\
    MB4              & 11 & $N+17$ & 25 \\
    MB4$^{\Gamma^0}$ & 4  & $N+10$ & 18 \\
    \bottomrule
  \end{tabular}
\end{table}

\begin{table*}[ht!]
\footnotesize
\centering
\caption{The geometric fingerprint features used in quantum machine learning for the five test molecules.}\label{tab:FG}
\begin{tabular}{lll}
\toprule
Molecule & Features & Width \\
\midrule
HF, N$_2$, CO
&
$\left[r_{12},\,r_{12}^{-1}\right]$
& 2 \\[2mm]

H$_2$O
&
$\left[
r_{\text{OH}_1},\,r_{\text{OH}_2},\,r_{\text{H}_1\text{H}_2},
\,r_{\text{OH}_1}^{-1},\,r_{\text{OH}_2}^{-1},
\,\angle_{\text{H}_1\text{OH}_2}
\right]$
& 6 \\[2mm]

NH$_3$
&
$\left[
r_{\text{NH}_1},\,r_{\text{NH}_2},\,r_{\text{NH}_3},
\,r_{\text{NH}_1}^{-1},\,r_{\text{NH}_2}^{-1},\,r_{\text{NH}_3}^{-1},
\,\angle_{\text{H}_1\text{NH}_2}
\right]$
& 7 \\[2mm]

\bottomrule
\end{tabular}
\end{table*}

\subsubsection{Transferability study}\label{sec:transferability}

To assess the transferability and extrapolation capabilities of the many-body encoding QML models, we apply the trained quantum circuits to predict electron-correlation energies at bond lengths beyond the training region. The testing range is chosen to encompass the nuclear displacements associated with the second vibrationally excited state ($v=2$) on the Hartree--Fock potential energy surface. 
Specifically, ten testing geometries are selected on each side of the training range, corresponding to bond compression and elongation, respectively. 
This provides a physically motivated benchmark for evaluating whether the QML models can generalize beyond the training data to molecular geometries relevant to vibrational dynamics. 

\subsubsection{Fingerprint baseline}

Fingerprint representations describe a molecule using a fixed-length
vector of precomputed features, such as interatomic distances,
bond angles, or connectivity patterns
\cite{Lilienfeld12_058301,Behler11_074106,Hahn10_742}.
Each feature is encoded as a single-qubit rotation angle
\cite{Killoran19_040504,Gambetta19_209}, without explicitly assigning
pair or triple descriptors to corresponding multiqubit gates.

The geometric fingerprint (FG) comprises interatomic distances,
their inverses, and bond angles, testing the predictive value of geometric information (see \cref{tab:FG}). For the FG baseline model, features are standardized using
training-set means and standard deviations and encoded through
single-qubit $R_Y$ rotations.

\begin{equation}
    \theta_i^{\mathrm{FG}}
=
\operatorname{clip}\left(
\frac{\pi}{2}
\frac{x_i-\mu_i}{\sigma_i},
-\pi,\pi
\right)
\end{equation}

\section*{Acknowledgments}
This exploratory work is supported by the U.S. Department of Energy, Office of Science, Basic Energy Sciences, in the Computational and Theoretical Chemistry program (Grant No. DE-SC0006863). The development of computational methods was supported by the National Science Foundation in the Chemical Theory, Models, and Computational Methods Program (CHE-2515209).

\section*{Code availability}

The code used to generate the descriptors, build the circuits, and train
the models will be made available at \url{https://github.com/novazhang25/Electronic-Structure-Informed-Many-Body-Encoding-QML}.

\section*{Author contributions}
L. Zhang and X. Li conceived the project. L. Zhu and L. Zhang developed the computer code. X. Li acquired computing resources. L. Zhang, Q. Yin, and L. Zhu carried out the calculations and analyzed the data. T. Culpitt provided the theoretical analysis. X. Li acquired research funding. X. Li, L. Zhu, L. Zhang, and T. Culpitt wrote the manuscript with input from all authors.  All authors discussed the results and approved the final version of the manuscript.

\section*{Competing interests}
The authors declare that there are no competing financial interests, personal relationships, or professional affiliations that could be perceived as influencing the work described in this manuscript.

\clearpage
\pagebreak

\bibliographystyle{unsrtnat}
\bibliography{Journal_Long_Name,Li_Group_References,refs}

\end{document}